\documentclass[12pt]{article}
\newcommand\mysection{\setcounter{equation}{0}\section}
\renewcommand{\theequation}{\thesection.\arabic{equation}}
\def\MSbar{\relax\ifmmode\overline{\rm MS}\else{$\overline{\rm MS}${ }}\fi}

\def\fun#1#2{\lower3.6pt\vbox{\baselineskip0pt\lineskip.9pt
  \ialign{$\mathsurround=0pt#1\hfil##\hfil$\crcr#2\crcr\sim\crcr}}}

\def\eV{{\rm e\kern-0.12em V}}            
  
\def\half{{\textstyle {1\over2}}}
  \def\quart{{\textstyle {1\over4}}}

\def\VEV#1{\left\langle #1\right\rangle}
\def \al {\relax\ifmmode{\alpha}\else{$\alpha${ }}\fi}
\def \be {\relax\ifmmode{\beta}\else{$\beta${ }}\fi}
\def\Im{\mathop{\rm Im}}    \def\Re{\mathop{\rm Re}}

\def\beq{\begin{equation}}   \def\eeq{\end{equation}}
\def \as{\relax\ifmmode\alpha_s\else{$\alpha_s${ }}\fi}
\def \pt{\relax\ifmmode{p_t}\else{$p_t${ }}\fi}

\def\eps{\relax\ifmmode\epsilon\else{$\epsilon${ }}\fi}
\def\ee{\relax\ifmmode{e^+e^-}\else{${e^+e^-}$}\fi}
\def\qq{\relax\ifmmode{q\overline{q}}\else{$q\overline{q}${ }}\fi}
\def\Np{\left(\begin{array}{c} N\\p \end{array}\right)}

\newskip\humongous \humongous=0pt plus 1000pt minus 1000pt
\def\caja{\mathsurround=0pt}
\def\eqalign#1{\,\vcenter{\openup1\jot
\caja   \ialign{\strut \hfil$\displaystyle{##}$&$
\displaystyle{{}##}$\hfil\crcr#1\crcr}}\,}

\newif\ifdtup

\def\eqal2#1{\,\vcenter{\openup1\jot
\caja   \ialign{\strut \hfil$\displaystyle{##}$&\hfil$
\displaystyle{{}##}$\hfil &$
\displaystyle{{}##}$\hfil\crcr#1\crcr}}\,}

\def\ib#1#2#3{{\em ibid.\ }~\underline{#1} (19#3) #2}
\def\np#1#2#3{{\em Nucl.\ Phys.\ }~\underline{B#1} (19#3) #2}
\def\pl#1#2#3{{\em Phys.\ Lett.\ }~\underline{#1B} (19#3) #2}
\def\pr#1#2#3{{\em Phys.\ Rev.\ }~\underline{D#1} (19#3) #2}

\def\prl#1#2#3{{\em Phys.\ Rev.\ Lett.\ }~\underline{#1} (19#3) #2}

 \def\cite#1{[\ref{#1}]}
 \def\citd#1#2{[\ref{#1},\ref{#2}]}
 
 \def\citq#1#2#3#4{[\ref{#1},\ref{#2},\ref{#3},\ref{#4}]}
 \def\citm#1#2{[\ref{#1}--\ref{#2}]}

\def\cF{{\cal{F}}}
\def\cM{{\cal{M}}}
\def\at{\al_{\mbox{\scriptsize eff}}}
\def\dat{\delta\at}
\def\re#1{(\ref{#1})}

\begin{document}
\begin{titlepage}
\begin{flushright}
CERN--TH/96-290\\
Cavendish--HEP--96/19\\
hep-ph/9612353
\end{flushright}              
\vspace*{\fill}
\begin{center}
{\Large \bf\boldmath Non-Perturbative Corrections to Heavy Quark\\[2mm]
Fragmentation in \ee\ Annihilation\footnote{Research supported in
part by the U.K. Particle Physics and Astronomy Research Council and by
the EC Programme ``Training and Mobility of Researchers", Network
``Hadronic Physics with High Energy Electromagnetic Probes", contract
ERB FMRX-CT96-0008.}}
\end{center}
\par \vskip 5mm
\begin{center}
        P.\ Nason$^{(1)}$ and B.R.\ Webber$^{(2)}$ \\
  $^{(1)}$Theory Division, CERN, CH-1211 Geneva 23, Switzerland\\    
  $^{(2)}$Cavendish Laboratory, University of Cambridge,\\
        Madingley Road, Cambridge CB3 0HE, U.K.
\end{center}
\par \vskip 2mm
\begin{center} {\large \bf Abstract} \end{center}
\begin{quote}
We estimate the non-perturbative power-suppressed corrections to heavy
flavour fragmentation and correlation functions in \ee\ annihilation, using
a model based on the analysis of one-loop Feynman graphs
containing a massive gluon. This approach corresponds to the
study of infrared renormalons in the large-$n_f$ limit of QCD,
or to the assumption of an infrared-finite effective
coupling at low scales.  We find that the leading
corrections to the heavy quark fragmentation function are of order $\lambda/M$,
where $\lambda$ is a typical hadronic scale
($\lambda\sim 0.4$ GeV) and
$M$ is the heavy quark mass. The inclusion of higher corrections
corresponds to convolution
with a universal function of $M(1-x)$
concentrated at values of its argument of order $\lambda$,
in agreement with intuitive expectations.
On the other hand, corrections to heavy quark correlations are very small,
of the order of $(\lambda/Q)^p$, 
where $Q$ is the centre-of-mass energy and $p \ge 2$.
\end{quote}
\vspace*{\fill}
\begin{flushleft}
     CERN--TH/96-290\\
     November 1996
\end{flushleft}
\end{titlepage}

\mysection{Introduction}

The production of hadrons containing heavy quarks in
\ee\ annihilation has proved to be a process of great
importance for testing the Standard Model and searching
for new physics. Heavy flavour processes are also valuable
as a testing-ground for new techniques in QCD, because the
heavy quark mass $M$ provides a second large momentum scale,
in addition to the overall hard scale $Q$, set by
the centre-of-mass energy in \ee\ annihilation. This has led
to a rather good understanding of heavy quark production and
fragmentation within the context of perturbation
theory \citm{MeNa}{NaOl}. On the other hand, it has become
clear as a result of this understanding that the $c$ and $b$
quark masses are not large enough for a purely perturbative
approach to provide a good description of the data for these
flavours \citd{CoNa}{DoKhTr}.  This is because non-perturbative
effects can give rise to contributions of order $\lambda/M$, where
$\lambda$ is a typical soft physics scale.
One may also worry about the possibility of
non-perturbative effects of order $\lambda/Q$,
which could be significant at present energies. Thus it
becomes important to study power-suppressed corrections
in QCD from as many viewpoints as possible, and to apply
the resulting insight to heavy flavour processes in particular.

One approach to the study of power-suppressed corrections,
which has proved popular recently, arose from the
study of infrared renormalons \cite{renormalons}.
Here a divergent series of perturbative
contributions gives rise to a power-suppressed renormalon
ambiguity in the prediction of perturbation theory.
One can then argue that a non-perturbative contribution
with the same power behaviour should be present, with an ambiguity
in its coefficient which cancels that associated with the renormalon.
Conversely, when the renormalon ambiguity involves a high inverse
power of the hard scale, one expects non-perturbative corrections
to be especially small. This approach can be reformulated without
reference to renormalons by postulating the existence of an
infrared-regular effective coupling at low scales (the
`dispersive approach' of ref.~\cite{DoMaWe}). The expected
power-suppressed corrections are consistent with those predicted
by more rigorous approaches such as the operator product expansion
where applicable, and have been found to agree fairly well with those
suggested by experimental data \citm{tube}{Stein}.

The aim of the present paper is to study power suppressed effects
in heavy flavour production in \ee\ annihilation, using either the
`renormalon' or `dispersive'
approach, which are equivalent for the purpose of the present work.
Our aim will be to 
test some standard assumption about the
form of non-perturbative corrections that enter the fragmentation
function. Furthermore, we will also examine the correlation between the
quark and antiquark momenta. This topic is of practical interest,
since this correlation affects the determination of $\Gamma_b$,
the partial width for the decay of the $Z^0$ boson into $b$-flavoured
hadrons.

\subsection{Renormalons and power-suppressed corrections}
In the renormalon approach, described in detail for \ee\ shape
variable calculations in ref.~\cite{NaSe}, one starts from the
first-order perturbative contributions to the process $\ee\to Q\bar Q X$,
involving emission of a single real or virtual gluon.  One then
identifies a factorially divergent series of perturbative
contributions associated with light quark loop insertions on
the gluon line. These contributions will be dominant at
high orders for sufficiently large values of $n_f$, the
number of light flavours. One now argues \citm{BBB}{LoMa}
that the true high-order behaviour of perturbative
QCD can be approximated
by making the replacement $n_f\to n_f -33/2$ in the
large-$n_f$ behaviour. There is support for this
`naive non-Abelianization' assumption in the high-order
behaviour of the average plaquette in quenched ($n_f=0$)
lattice QCD \cite{DiOnMa}.

The resulting factorial divergences of the perturbation series
can be of two types.  Those associated with high-momentum
regions of integration (ultraviolet renormalons) correspond
to divergent series with alternating signs, which can be summed
unambiguously using standard techniques such as Borel summation.
We shall not be concerned with them in the present paper.
Those due to low momenta flowing in loop integrals
(infrared renormalons) produce same-sign asymptotic
series, which are intrinsically ambiguous. The ambiguity
is of the order of the smallest term in the series, which
turns out to be a power-suppressed quantity,
of order $(\lambda/Q)^p$ where $p$ is
a number that depends on the observable being computed.

In the full theory of QCD, the infrared renormalon
ambiguities of perturbation theory must be cancelled
by non-perturbative contributions. We shall assume
that the presence of an infrared renormalon with a
particular value of the power $p$ indicates that
a comparable power-suppressed non-perturbative
contribution is actually present in the full theory.

The dispersive approach of ref.~\cite{DoMaWe} involves
similar calculations, with a slightly different
interpretation. A (formal) dispersion relation for the
QCD running coupling of the form
\beq\label{nonAb-disrel}
\as(k^2) = - \int_0^\infty \frac{d\mu^2}{\mu^2+k^2} \>\rho_s(\mu^2)
\,, \;\;\;\;\;\;\;
\rho_s(\mu^2) =
-\frac{1}{2\pi i}\, \mbox{Disc} \left\{\as(-\mu^2)\right\} \,,
\eeq
is assumed. In QCD, unlike QED, the running of
the coupling cannot be associated with vacuum polarization
effects alone. However, in the same spirit as the
`naive non-Abelianization' assumption,
it is further assumed that the dominant
effect on some QCD observable $F$ of the running of $\as$
in one-loop graphs may be represented in terms of the
spectral function $\rho_s(\mu^2)$ and a characteristic
function $\cF(\mu^2)$, as follows:
\beq\label{WDM1}\eqalign{
F &= \as(0)\cF(0) + \int_0^\infty 
\frac{d\mu^2}{\mu^2}\rho_s(\mu^2)
\cdot\cF(\mu^2)\cr
&= \int_0^\infty \frac{d\mu^2}{\mu^2}\rho_s(\mu^2)
\cdot
\left[\,\cF(\mu^2) - \cF(0)\,\right]\;,}
\eeq
where the relation \re{nonAb-disrel} has been used 
to eliminate $ \as(0)$. The characteristic function $\cF(\mu^2)$
is obtained by computing the relevant one-loop graphs with a
non-zero gluon mass $\mu$ \citd{BBB}{Web94} and dividing
by $\as$.
Integrating Eq.~\re{WDM1} by parts, we can write
\beq\label{WDMF}
F\>=\>  \int_0^\infty
\frac{d\mu^2}{\mu^2}\> \at(\mu^2) \cdot \dot\cF (\mu^2) \>,
\eeq
where
\beq
\dot{\cF} \equiv -\frac{\partial\cF}{\partial\ln \mu^2}
\eeq
and we have introduced the effective coupling $\at(\mu^2)$,
defined in terms of $\rho_s(\mu^2)$ by
\beq\label{aeffrel}
 \rho_s(\mu^2) =\frac{d}{d\ln\mu^2} \at(\mu^2) \>.
\eeq
It follows from this definition and Eq.~\re{nonAb-disrel} that
\beq\label{atexpn}
 \at(\mu^2)\>=\> \as(\mu^2) -\frac{\pi^2}{3!}\frac{d^2\as}{d\ln^2\mu^2}
 +\frac{\pi^4}{5!}\frac{d^4\as}{d\ln^4\mu^2}
 - \ldots \>=\> \as\,+\, {\cal O}(\as^3)\>,
\eeq
and therefore in the perturbative domain $\as\ll 1$ the standard and
effective couplings are approximately the same.
In the large-$n_f$ renormalon
approach\footnote{We stress that this model is not physically fully
consistent, because of the presence of the Landau pole,
which implies that the support of the
spectral function of $\as$ must be extended to negative values
of the argument for Eq.~\re{nonAb-disrel} to be valid.}
we have the explicit expression,
obtained by substituting the one-loop running coupling
into Eq.~\re{atexpn}:
\beq
\at(\mu^2) = \frac{1}{\pi b_0}\arctan\left(\pi b_0\as(\mu^2)\right)
\eeq
where $b_0=(33-2 n_f)/12\pi$.

The characteristic function $\cF(\mu^2)$ is more precisely a function
of the dimensionless ratio $\eps = \mu^2/Q^2$, where $Q^2$ is the
characteristic scale of the hard process. If the effective
coupling in Eq.~\re{WDMF} has a non-perturbative component
$\dat(\mu^2)$, with support limited to low values of $\mu^2$,
the corresponding correction to $F$,
\beq\label{Fxxbar}
\delta F = \int\frac{d\mu^2}{\mu^2}\dat(\mu^2)\,\dot\cF(\mu^2)\;,
\eeq
will therefore have a $Q^2$ dependence determined by the low-$\mu^2$
behaviour of $\cF$.

A crucial point is that only those terms in $\cF$ that are
non-analytic at $\mu^2=0$ ($\eps=0$) can produce power-suppressed
contributions to $\delta F$. This is because the integer
$\mu^2$-moments of $\dat(\mu^2)$ are required to vanish, for
consistency with the operator product expansion.  The same result may
be seen in the renormalon analysis of ref.~\cite{NaSe}: for any
behaviour of $\cF(\eps)$ of the form $\eps^p$ as $\eps\to 0$, the
renormalon contribution is proportional to $\Im(e^{ip\pi})$ and
therefore vanishes for integer $p$.  On the other hand, a
$\sqrt{\eps}$ behaviour implies a non-vanishing correction
proportional to $1/Q$, while $\eps\ln\eps$ gives $1/Q^2$, etc. Thus
our objective is to identify the leading non-analytic terms in the
behaviour of the characteristic function at small values of the gluon
mass-squared, which will tell us the $Q^2$-dependence of the leading
power-suppressed corrections.

If one makes the additional assumption that the effective
coupling modification $\dat(\mu^2)$ in Eq.~\re{Fxxbar}
is universal, one obtains a factorization property for
power-suppressed corrections, which leads to
relationships between the coefficients of the
corrections to different observables.
For variables like event shapes, this type of factorization
is only approximate, due to the fact that a cut dressed gluon line
is weighted differently, depending upon the value of the shape
variable for the particular final-state structure
of the cut gluon \cite{NaSe}.
In the present case, however, the dressed gluon is cut fully inclusively,
without any weight, and therefore factorization may be more reliable.

In the case of heavy flavour processes, we also want to
study corrections that are suppressed by powers of the
heavy quark mass, $M$. As long as we treat both $M$ and $Q$
as large parameters, and keep track of the dependence on their
ratio, this will be done automatically when we extract the
non-analytic terms in $\eps$. Defining $\rho = 4M^2/Q^2$,
a $\sqrt{\eps/\rho}$ term will indicate a correction
proportional to $1/M$, $\eps\ln\eps/\rho$ implies $1/M^2$,
and so on.

In our terminology, mass corrections of the form $(M/Q)^p$
will not be called power-suppressed, since we are always assuming
that $M$ is not small.

\mysection{Calculations}
\subsection{Massive gluon cross sections}

Considering first the vector current contribution,
the distribution of the heavy quark and antiquark energy fractions
$x$ and $\bar x$ with emission of a gluon of mass $\mu$ in the
process $\ee\to Q\bar Q g$ is given by
\beq\label{dsigdxdxb}
\frac{1}{\sigma_V}\frac{d^2\sigma_V}{dx\,d\bar x} =
\frac{\as}{2\pi}\frac{C_F}{\beta}
\left[ \frac{(x+\eta)^2+(\bar x+\eta)^2+\zeta_V}
{(1+\half\rho)(1-x)(1-\bar x)} -\frac{\eta}{(1-x)^2}
-\frac{\eta}{(1-\bar x)^2}\right]\;.
\eeq
Here $\eta = (\mu^2+2M^2)/Q^2 = \eps +\half\rho$ where
$\eps=\mu^2/Q^2$, $M$ is the quark mass, $\rho = 4 M^2/Q^2$,
\beq\label{zetaV}
\zeta_V = -2\rho(1+\eta)\;,
\eeq
$\beta =\sqrt{1-\rho}$ is the heavy quark velocity, and
\beq\label{sigmaQ}
\sigma_V = \sigma_0\left(1+\half\rho\right)\beta
\eeq
is the Born cross section for heavy quark production by a vector
current, $\sigma_0$ being the massless quark Born cross section.

The phase space is determined by the triangle relation
\beq\label{triangle}
\Delta(x^2-\rho,\bar x^2-\rho,x_g^2-4\eps)\leq 0
\eeq
where $\Delta(a,b,c) = a^2+b^2+c^2-2ab-2bc-2ca$
and $x_g = 2-x-\bar x$. This gives $x_-\leq \bar x\leq x_+$
where
\beq\label{xpm}\eqalign{
& x_\pm \>=\> \frac{(2-x)(1-x-\eps+\half\rho)\pm \Xi(x,\rho,\eps)}
{2(1-x)+\half\rho}\cr
& \Xi(x,\rho,\eps) \>=\> \sqrt{(x^2-\rho)[(1-x-\eps)^2-\eps\rho]}\;,}
\eeq
and
\beq
\sqrt{\rho}\leq x \leq 1-\eps-\sqrt{\eps\rho}\;.
\eeq
In the case of the axial current contribution, instead
of Eq.~\re{dsigdxdxb} we have
\beq\label{dsigA}
\frac{1}{\sigma_A}\frac{d^2\sigma_A}{dx\,d\bar x} =
\frac{\as}{2\pi}\frac{C_F}{\beta}
\left[\frac{(x+\eta)^2+(\bar x+\eta)^2+\zeta_A}
{(1-\rho)(1-x)(1-\bar x)}
-\frac{\eta}{(1-x)^2}
-\frac{\eta}{(1-\bar x)^2}\right]\;,
\eeq
where
\beq\label{zetaA}
\zeta_A = \half\rho[(3+x_g)^2-19+\rho-8\eps]\;,
\eeq
$\sigma_A$ being the Born cross section for heavy quark
production by the axial current:
\beq\label{sigmaA}
\sigma_A = \sigma_0\beta^3 = \sigma_0 (1-\rho)\beta\;.
\eeq

\subsection{Leading power corrections}

Clearly the expressions \re{dsigdxdxb} and \re{dsigA}
are analytic functions of $\eps$ at $\eps=0$
except possibly for $x$ and $\bar x$ near 1.
The phase space is also analytic in $\eps$ whenever the gluon
momentum is large, since in this region one can always expand
kinematic variables in powers of $\eps$.
As discussed above, this implies that there are no non-perturbative
corrections of the type we are considering for $x,\bar x <1$.
Non-analytic behaviour may only arise in the region where
$x_g \approx\sqrt{\eps}$ ($x,\bar x\approx 1$).
In order to investigate the corrections associated with this
region we take moments of the form
\beq
\cM(N,\bar N,\eps) = 
\int dx d\bar x x^N \bar x^{\bar N}
\frac{1}{\sigma}\frac{d^2\sigma}{dx\,d\bar x}
\eeq
and expand
\beq
x^N \bar x^{\bar N} = 1- Ny -\bar N\bar y +\cdots\;,
\eeq
where $y=1-x$ and $\bar y=1-\bar x$. The first term corresponds to
the total heavy flavour cross section, whose dominant
power correction is 
of order $(\lambda/Q)^4$ or smaller.  However, the next two
terms give contributions proportional to $\sqrt{\eps}$ at
small $\eps$, which could give rise to $\lambda/Q$ and/or
$\lambda/M$ corrections. To
evaluate them we note that their contribution to the difference
$\delta\cM =\cM(N,\bar N,0)- \cM(N,\bar N,\eps)$ for small
$\eps$ can be written, for both the vector and axial
current contributions, as $C_F\as\cF/2\pi$ where
\beq\label{deltaF}\eqalign{
\cF &= \frac{1}{\beta}\int_R dy d\bar y (Ny +\bar N\bar y)
\left(\frac{2-\rho}{y\bar y} -\frac{\rho}{2y^2}
-\frac{\rho}{2\bar y^2}\right)\cr
&= \frac{1}{2\beta}(N+\bar N)\int_R dy d\bar y 
\left(\frac{4-3\rho}{y} -\frac{\rho\bar y}{y^2}\right)\;,}
\eeq
$R$ being the region between the phase space boundaries
for $\eps=0$ and $\eps>0$.  Eq.~\re{triangle} may be expanded
in the region $x_g=y+{\bar y}\approx \sqrt{\eps}$, so that
the boundary of phase space in this region is given by
\beq\label{triangleapprox}
\quart\Delta(x^2-\rho,\bar x^2-\rho,x_g^2-4\eps)\approx
\rho(y^2+{\bar y}^2)-2(2-\rho)y{\bar y}+4\eps(1-\rho)\leq 0\,,
\eeq
which is the equation of a hyperbola.
Changing variables to $r,\phi$ where $y=r\cos\phi$ and
$\bar y=r\sin\phi$, the region $R$ may be written as
\beq
R:\;\;\;\;\delta<\phi<\half\pi-\delta\;,\;\;
0<r<\sqrt{\frac{2\eps(1-\sin 2\delta)}{\sin 2\phi-\sin 2\delta}}
\eeq
where $\sin 2\delta = \rho/(2-\rho)$.
Performing the integration one gets
\beq\label{FNNbar}\eqalign{
\cF &= \frac{\pi}{2\beta}(N+\bar N)
\sqrt{\frac{2\eps(1-\sin 2\delta)}{\sin 2\delta}}
\left(4-3\rho -\frac{\rho}{\sin 2\delta}\right)\cr
&=2\pi\beta^2 (N+\bar N)\sqrt{\frac{\eps}{\rho}}
= \pi (N+\bar N)\left(1-4\frac{M^2}{Q^2}\right)
\frac{\mu}{M}\;.}
\eeq
Therefore the
leading power correction is of order $\mu/M$
rather than $\mu/Q$. The leading correction with an explicit
dependence on $Q$ is of order $\mu M/Q^2$. This is
consistent with the finding for light quark fragmentation
functions \citq{DoMaWe}{DaWefra}{BeBrMa}{BaBr}: the
leading power correction in $Q$ is of order $1/Q^2$.

The linear dependence on the moment index $N$ in the
result \re{FNNbar} implies a behaviour in $x$-space of the form
\beq\label{cFxxbar}
\cF(x,\bar x,\mu^2) =\pi\beta^2 \frac{\mu}{M}
\left[\delta(1-x)\delta'(1-\bar x)+\delta'(1-x)\delta(1-\bar x)\right]\;.
\eeq
This means that, as far as the leading power correction
is concerned, the two-particle heavy quark distribution
factorizes. In the dispersive approach of ref.~\cite{DoMaWe},
the non-perturbative correction is given in terms of the low-energy
modification to the effective coupling, $\dat(\mu^2)$, by
Eq.~\re{Fxxbar}. Defining the non-perturbative parameter
\beq
A_1 = \frac{C_F}{2\pi}\int\frac{d\mu^2}{\mu^2}\,\mu\,\dat(\mu^2)\;,
\eeq
Eqs.~\re{Fxxbar} and \re{cFxxbar} then imply that
\beq\label{dsdxdxbar}\eqalign{
\frac{1}{\sigma}\frac{d^2\sigma}{dx\,d\bar x} &\approx
\left[\delta(1-x)-\frac{\pi A_1}{2M}\beta^2\delta'(1-x)\right]\cdot
\left[\delta(1-\bar x)-\frac{\pi A_1}{2M}\beta^2
\delta'(1-\bar x)\right]\cr
&\approx \delta\left(1-x-\beta^2\frac{\lambda}{M}\right)
\cdot\delta\left(1-\bar x-\beta^2\frac{\lambda}{M}\right)}
\eeq
where $\lambda = \pi A_1/2$.
Thus we see that the main non-perturbative effect is a shift in
the heavy-quark momentum fractions by an amount $\delta x\sim\lambda/M$.
Assuming approximate universality of $\dat$, one may estimate
from light-quark event shape data that $A_1\approx 0.25$ GeV
\cite{DoMaWe}, which gives $\lambda\sim 0.4$ GeV. This agrees
with the order of magnitude of the non-perturbative
shift estimated from $\VEV{x}$ in heavy flavour
fragmentation \cite{CoNa}.

\subsection{Higher power corrections}

Power-suppressed effects
in the heavy flavour fragmentation functions should be equivalent
to a convolution with a non-perturbative initial condition
of the form $M f\left(M(1-x)\right)$, and therefore should approach a delta
function as $M\to\infty$.
This can be inferred by
intuitive reasoning, but can also be derived more rigorously
in the context of the heavy quark mass expansion \cite{JaRa}.
In this section we will show that this
expectation is also fulfilled in our model.

First of all we note that Eq.~\re{dsdxdxbar} can be
written as
\beq\label{dsdkdkbar}
\frac{1}{\sigma}\frac{d^2\sigma}{dx\,d\bar x}
\approx M\delta\left(M(1-x)-\beta^2\lambda\right)
\cdot M\delta\left(M(1- \bar x)-\beta^2\lambda\right)
\eeq
and therefore the expected form is indeed obtained when
one includes only the leading power correction.

To go beyond the leading correction, we have to
consider higher moments with respect to $y$ and $\bar y$ in
Eq.~\re{deltaF}. We examine first the moments of the
single-particle distribution ($\bar N=0$). For moments
weighted by $y^p$ with $p>1$ we have to define the integration
region $R$ more carefully. Consider for simplicity the case
that $\rho$ is small (i.e.\ $M^2\ll Q^2$), so that
$\sin 2\delta\approx 2\delta\approx\half\rho$.
The upper limit of the $r$-integration becomes
a constant of order unity when $\delta<\phi <\delta+\eps$.
Hence this region gives a term that is analytic in
$\eps$, which will not contribute to power corrections.
The important region is $\delta+\eps<\phi <\half\pi-\delta-\eps$.
The leading non-analytic term coming from this region
is proportional to $(\eps/\delta)^{p/2}$ when $p$ is
odd, and proportional to $(\eps/\delta)^{p/2}\ln{\eps}$ when
$p$ is even. Hence for every value of $p$ there is a power
correction of order $(\eps/\delta)^{p/2}= (\mu/M)^p$.

In detail, for $\rho$ small and $\bar N =0$ Eq.~\re{deltaF}
becomes
\beq\label{deltaFN}\eqalign{
\cF &= \sum_{p=1}^\infty\Np\int_R dy d\bar y\, y^p
\left(\frac{2}{y\bar y} -\frac{\rho}{2y^2}
-\frac{\rho}{2\bar y^2}\right)\cr
&\approx \sum_{p=1}^\infty
\Np\frac{2}{p}\,\eps^{p/2}\int_{\eps+\delta}^1
\frac{d\phi}{\phi^2}\,(\phi-\delta)^{(2-p)/2}\;.}
\eeq
Now
\beq\label{inteps}\eqalign{
\int_{\eps+\delta}^1
\frac{d\phi}{\phi^2} (\phi-\delta)^{(2-p)/2}
&\approx (-1)^{(p-1)/2}\left(1-\half p\right)\pi\delta^{-p/2} +\cdots
\;\;\;\;\mbox{for $p$ odd,}\cr
&\approx (-1)^{p/2}\left(1-\half p\right)\delta^{-p/2}\ln\eps +\cdots
\;\;\;\;\;\;\mbox{for $p$ even,}}
\eeq
where the dots correspond to terms giving contributions that
are analytic and/or higher-order in $\eps$. Thus, keeping only
the leading non-analytic parts, we find
\beq\label{deltaFsum}
\cF = \Re\left\{(\ln\eps-i\pi)
\sum_{p=1}^\infty\Np\left(\frac{2}{p}-1\right)
\left(\frac{i\mu}{M}\right)^p\right\}\;,
\eeq
which corresponds in $x$-space to
\beq\label{deltaFx}\eqalign{
\cF(x,\mu^2) &= M f\left(M(1-x) \right) \cr
f(z)&=
 \Re\left\{(\ln\eps-i\pi)
\sum_{p=1}^\infty\frac{1}{p!}\left(\frac{2}{p}-1\right)
(i\mu)^p\,\delta^{(p)}(z)\right\}\;.}
\eeq
We can therefore see the expected scaling of the fragmentation function.
We also see that while the leading power correction, eq.~\re{dsdxdxbar},
corresponds to a simple shift in the value of $x$,
this behaviour is not preserved by the higher power corrections.
%

\mysection{Correlations}
We consider now the higher power corrections to the
two-particle distribution, i.e.\ the inclusion of higher
powers of $\mu/M$ in Eq.~\re{cFxxbar}. We have to
examine the double moments corresponding to Eq.~\re{deltaF}
with general weights $y^p\bar y^q$. Again we treat
only the case of small $\rho$.
Then we find that for any $q$ such that $0<q\le p$, the
leading non-analytic term is suppressed by a factor of $\rho^q$
relative to that for weight $y^{p+q}$.  Therefore the leading
terms in each order of $\mu/M$ remain of the form \re{cFxxbar}:
\beq
\cF(x,\bar x,\mu^2) =
\delta(1-x)\cF(\bar x,\mu^2)+\cF(x,\mu^2)\delta(1-\bar x)
\eeq
where $\cF(x,\mu^2)$ is as given in Eq.~\re{deltaFx}.
Thus the two-particle distribution including these terms
still factorizes and can be expressed as a function of
$M(1-x)$ and $M(1-\bar x)$, although
beyond leading order in $1/M$ it differs from the simple
product of delta functions given in Eq.~\re{dsdkdkbar}.

Because of its possible impact on the determination of $\Gamma_b$
in $Z^0$ decays,
it is interesting to determine what is the leading power correction
to the momentum correlation $\langle y{\bar y}\rangle-\langle y \rangle^2$.
As stated before, the correction to $\langle y{\bar y}\rangle$
behaves as $\rho(\sqrt{\eps/\rho})^2=\eps$ at small $\eps$.
The term $\langle y \rangle^2$ gives
zero at the order we are considering. In fact in the large-$n_f$ limit
it is subleading, and in the dispersive approach it is
of second order in the effective coupling. Therefore the
momentum correlation is of order $\eps$.

In order to confirm this conclusion, we also calculated
the difference $\langle y{\bar y}\rangle-\langle y{\bar y}\rangle_{\eps=0}$
using the exact phase space and matrix elements. We found
that the leading term at small $\eps$ is
proportional to $\eps\ln\rho$.
Thus, corrections to the momentum correlation
in our model are suppressed by at least
two powers of $Q$, and should
therefore be completely negligible at LEP energies. Whether this result
survives higher-order corrections is an open question, and in fact a very
difficult one. We simply point out here that, while Monte Carlo models
seem to indicate the presence of $\lambda/Q$ corrections to
correlations (see ref.~\cite{NaOl} and references therein),
the simple model that we have adopted in this
work does not provide support for the presence of such corrections.
This is also consistent with the findings of ref.~\cite{DaWefra}, where
power corrections to fragmentation functions were computed in the strictly
massless limit.

\mysection{Discussion}

We have examined the heavy flavour fragmentation function and correlations
in a simple model, and found the following results.

At leading order, non-perturbative effects in the fragmentation function
can be represented as a convolution with a function of the form
\beq
M\; f\left(M(1-x)\right)\,,
\eeq
which approaches a $\delta$ function as $M\to\infty$.
The leading power correction has the form $\lambda/M$, where $\lambda$ is
a typical soft
hadronic scale. An estimate of this correction based on the approach
proposed in ref.~\cite{DoMaWe} gives the correct sign and order of
magnitude.

In the two-particle distribution, corrections of the order of
$(\lambda/M)^p$ factorize and therefore no large correlations
of this order arise.
Since correlations are important for their possible
impact on the determination of $\Gamma_b$ in $Z^0$ decays,
and since the perturbative value of the correlation is of the order of
1\% \cite{NaOl}, it is also important to understand whether
corrections of the order of $\lambda/Q$ are present.
In our analysis, consistently with ref.~\cite{DaWefra}, terms of this
order do not arise. In fact, we also verified numerically that
the $\eps$ dependence of the correlation $\langle y{\bar y}\rangle
-\langle y \rangle^2$ is of order $\eps \log\rho$, and therefore
the leading power correction is less than order $(\lambda/Q)^2\log\rho$.

We end with a comment on the relationship between our results and those
of ref.~\cite{DoKhTr}. In that paper a resummed expression for the
heavy quark spectrum was derived and numerical results were presented
using various models for the behaviour of the QCD running coupling
at low scales. The non-perturbative component of the coupling
generates $1/M$ corrections which should correspond to those considered
here, after convolution with the perturbative fragmentation
function.

\par \vskip 2ex\noindent{\bf Acknowledgements}\\ \noindent
We would like to thank Yu. Dokshitzer for useful comments.
BRW is grateful to the CERN Theory Division for hospitality
during part of this work. 

\section*{Appendix}
\setcounter{equation}{0}
\renewcommand{\theequation}{A.\arabic{equation}}

We give here for reference the single-particle inclusive distribution which
results from integrating the vector two-particle distribution \re{dsigdxdxb}
for $x<1$:
\beq\label{dsigdx}\eqalign{
\frac{1}{\sigma_V}\frac{d\sigma_V}{dx} &=
\int_{x_-}^{x_+}\frac{d\bar x}{\sigma_V}\frac{d^2\sigma_V}{dx\,d\bar x}\cr
&= \frac{\as}{2\pi}\frac{C_F}{\beta(1+\half\rho)}
\Biggl\{\left(\frac{2(1+\eps)^2-\half\rho^2}{1-x}-1-x-2\eps-\rho\right)\cr
&\times\ln\left[\frac{[(1-x)(x-\half\rho)+\eps(2-x)+\Xi(x,\rho,\eps)]^2}
{(1-x+\quart\rho)[\rho(1-x)^2+4\eps(x+\eps-\rho)]}\right]\cr
&-\frac{\Xi(x,\rho,\eps)\,\Phi(x,\rho,\eps)}
{8(1-x)^2(1-x+\quart\rho)^2[\rho(1-x)^2+4\eps(x+\eps-\rho)]}\Biggr\}}
\eeq
where $x_\pm$ and $\Xi$ are given in Eq.~\re{xpm} and
\beq\eqalign{
&\Phi(x,\rho,\eps) \>=\> \rho(1-x)^2[4(1-x)^2(8-x)
+2\rho(1-x)(17-9x)+2\rho^2(5-4x)+\rho^3]\cr
&+2\eps[8(1-x)^2(2+x^2)-2\rho(1-x)(4-7x-12x^2+7x^3)\cr
&\;\;\;\;\;\;\;\;-2\rho^2(13-19x+x^2+4x^3)
-\rho^3(9-7x-x^2)-\rho^4]\cr
&+2\eps^2[8(1-x)(4-x-2x^2)+4\rho(1+5x-5x^2)
-10\rho^2(1-x)-\rho^3]\cr
&+4\eps^3[4(1-x)(4-3x)+2\rho(5-4x)+\rho^2]\;.}
\eeq

\par \vskip 1ex
\noindent{\large\bf References}
\begin{enumerate}
\def\cav#1{Cambridge preprint Cavendish--HEP--#1}
\def\cern#1{CERN preprint TH/#1}
\item\label{MeNa}
       B.\ Mele and P.\ Nason, \pl{245}{635}{90}; \np{361}{626}{91}.
\item\label{CoNa}
       G.\ Colangelo and P.\ Nason, \pl{285}{167}{92}.
\item\label{DoKhTr}
       Yu.L.\ Dokshitzer, V.A.\ Khoze and S.I.\ Troyan, \pr{53}{89}{96}. 
\item\label{NaOl}
       P.\ Nason and C.\ Oleari, \pl{387}{623}{96}.
\item\label{renormalons}
       For reviews and classic references see:\\
       V.I. Zakharov, \np{385}{452}{92};\\ 
       A.H.\ Mueller, in {\em QCD 20 Years Later}, vol.~1
       (World Scientific, Singapore, 1993).
\item\label{DoMaWe}
       Yu.L.\ Dokshitzer, G.\ Marchesini and B.R.\ Webber,
       \np{469}{93}{96}.
\item\label{tube}
       B.R.\ Webber, in {\em Proc.\ Summer School on Hadronic Aspects of
       Collider Physics,\\
       Zuoz, Switzerland, 1994} [hep-ph/9411384].
\item\label{Web94}
       B.R.\ Webber, \pl{339}{148}{94};\\
       Yu.L.\ Dokshitzer and B.R.\ Webber, \ib{352B}{451}{95}.
\item\label{DaWeDIS}
       M.\ Dasgupta and B.R.\ Webber, \pl{382}{273}{96}.
\item\label{DaWefra}
       M.\ Dasgupta and B.R.\ Webber, \cav{96/9} [hep-ph/9608394],
       to be published in Nucl.\ Phys.\ B.
\item\label{BeBrMa}
       M.\ Beneke, V.M.\ Braun and L.\ Magnea, Stanford preprint
       SLAC-PUB-7274 [hep-ph/9609266]. 
\item\label{Stein}
       E.\ Stein, M.\ Meyer-Hermann, L.\ Mankiewicz and A.\ Sch\"afer,
       \pl{376}{177}{96};\\
       M.\ Meyer-Hermann, M.\ Maul, L.\ Mankiewicz,
       E.\ Stein and A.\ Sch\"afer, Frankfurt preprint UFTP-414-1996
       [hep-ph/9605229].
\item\label{NaSe}
       P.\ Nason and M.H.\ Seymour, \np{454}{291}{95}.
\item\label{BBB}
       M.\ Beneke, V.M.\ Braun and V.I.\ Zakharov, \prl{73}{3058}{94};\\
       P.\ Ball, M.\ Beneke and V.M.\ Braun, \np{452}{563}{95};\\
       M.\ Beneke and V.M.\ Braun, \np{454}{253}{95}.
\item\label{BroKa}
       D.J.\ Broadhurst and A.L.\ Kataev, \pl{315}{179}{93}.
\item\label{LoMa}
       C.N.\  Lovett-Turner and C.J.\ Maxwell, \np{452}{188}{95}. 
\item\label{DiOnMa}
       F.\ Di Renzo, E.\ Onofri and G.\ Marchesini, \np{457}{202}{95}.
\item\label{BaBr}
       I.I.\ Balitsky and V.M.\ Braun, \np{361}{93}{91}.
\item\label{JaRa}
       R.L. Jaffe and L. Randall, \np{412}{79}{94}. 
\end{enumerate}
\end{document}